\documentclass[aps,showpacs,reprint,prb,superscriptaddress]{revtex4-1}
\usepackage{graphics}
\usepackage{graphicx}
\usepackage{dcolumn}

\begin{document}

\title{Local structural disorder and superconductivity in K$_{x}$Fe$_{2-y}$Se%
$_{2}$}
\author{Hyejin Ryu}
\author{Hechang Lei}
\author{A. I. Frenkel}
\email{anatoly.frenkel@yu.edu}
\author{C. Petrovic}
\email{petrovic@bnl.gov}
\date{\today }

\begin{abstract}
We report significantly enhanced magnetic moment in K$_{0.69(2)}$Fe$%
_{1.45(1)}$Se$_{2.00(1)}$ single crystals with sharp T$_{c}$ and bulk
superconductivity obtained by post-annealing and quenching process. There
are two Fe sites in K$_{0.69(2)}$Fe$_{1.45(1)}$Se$_{2.00(1)}$ unit cell: Fe1
which has higher symmetry with longer average Fe-Se bond length, and Fe2
which has lower symmetry with shorter average Fe-Se bond length. Temperature
dependent X-ray absorption fine structure (XAFS) analysis results on
quenched and as-grown K$_{0.69(2)}$Fe$_{1.45(1)}$Se$_{2.00(1)}$ crystals
show that quenched K$_{0.69(2)}$Fe$_{1.45(1)}$Se$_{2.00(1)}$ have increased
average Fe-Se bond length and decreased static disorder. Our results
indicate that nonzero population of Fe1 sites is the key structural
parameter that governs the bulk superconductivity. We also show clear
evidence that Fe1 sites carry higher magnetic moment than Fe2 sites.
\end{abstract}

\maketitle

\affiliation{Condensed Matter Physics and Materials Science Department, Brookhaven
National Laboratory, Upton, NY 11973, USA}

\affiliation{Condensed Matter Physics and Materials Science Department, Brookhaven
National Laboratory, Upton, NY 11973, USA}

\affiliation{Department of Physics, Yeshiva University, New York, New York 10016,
USA}

\affiliation{Condensed Matter Physics and Materials Science Department, Brookhaven
National Laboratory, Upton, NY 11973, USA}

\section{Introduction}

The discovery of superconductivity in LaOFeAs$_{1-x}$F with transition
temperature T$_{c}$ up to 26 K stimulated a variety of study on iron based
superconductors.\cite{Kamihara} The T$_{c}$ in arsenides was soon raised up
to 55 K and was also discovered in simple binary structures of selenide
materials: FeSe,\cite{Hsu FC} FeTe$_{1-x}$Se$_{x}$,\cite{KWYeh} and FeTe$%
_{1-x}$S$_{x}$\cite{Mizuguchi} that do not have any crystallographic layer
in between puckered Fe-Se(Te) sheets. FeSe is easily affected by pressure
since the T$_{c}$ can be increased from 8 K to 37 K with dT$_{c}$/dP $\sim $
9.1 K/GPa. One possible explanation for this behavior is the empirical rule
that the T$_{c}$ correlates with the anion height between Fe and Se layers,
with an optimal distance around 0.138 nm for maximum T$_{c}$ $\sim $ 55 K.%
\cite{Mizuguchi2} Recently, the superconducting T$_{c}$ was enhanced to
above 30 K in iron selenide material not by external pressure but by
intercalating alkaline metal (K) between the FeSe layers (AFeSe-122 type).%
\cite{Guo} In addition, it is reported that the superconducting state can be
obtained from an insulating state by post-annealing and fast quenching.\cite%
{Han}

In this work, we have exploited Fe and Se K-edge spectra using X-ray
absorption fine structure (XAFS) of as-grown and quenched K$_{0.69(2)}$Fe$%
_{1.45(1)}$Se$_{2.00(1)}$ in order to examine the local lattice and
electronic structure around the Fe and Se atoms. We show strong evidence
that superconducting volume fraction increase is intimately connected with
the increased occupancy of the high symmetry Fe site, accompanied by the increased average Fe-Se distance and decreased average configurational (static) disorder in this distance.

\section{Experiment}

As grown and quenched K$_{0.69(2)}$Fe$_{1.45(1)}$Se$_{2.00(1)}$ single
crystals were prepared as described previously.\cite{Hechang1,Hechang2}
X-ray absorption experiments were completed at beamline X19A of the National
Synchrotron Light Source. Temperature-dependent X-ray absorption data were
collected in the transmission mode. Gas-filled ionization chamber detectors
were used for incident, transmitted, and reference channels. A closed cycle
He cryostat was used to cool the samples with temperature control within $%
\pm $1 K. A minimum of two scans were measured for each temperature for
optimal signal to noise ratio. All XAFS spectra were analyzed using the
Athena and Artemis software programs.\cite{Ravel}

We compared several different modeling schemes in fitting FEFF6 theory to the experimental data in order to obtain structural information from XAFS analysis. The models compared where: 1) the multiple edge model, where we varied Fe and Se edge data concurrently, by constraining their bond lengths and their disorders to be the same at each temperature, 2) the model where we added third cumulant to Fe-Se contribution of each data set, 3) the multiple data set model where we constrained the disorder parameter to follow the Einstein model with static disorder (vide infra), and 4) the model where we added Fe-Fe contribution to Fe edge fits. After comparing the fit qualities and inspecting the best fit results for their physical meaning, we chose the model (3) for presenting our results, although the main trends in the results remained the same across all models we tried. We found that adding third cumulant to the final fit model did not change the results within the error bars, and the best fit values of the third cumulant were consistent with zero.

\section{Results and Discussion}

\begin{figure}[tbp]
\centerline{\includegraphics[scale=1.05]{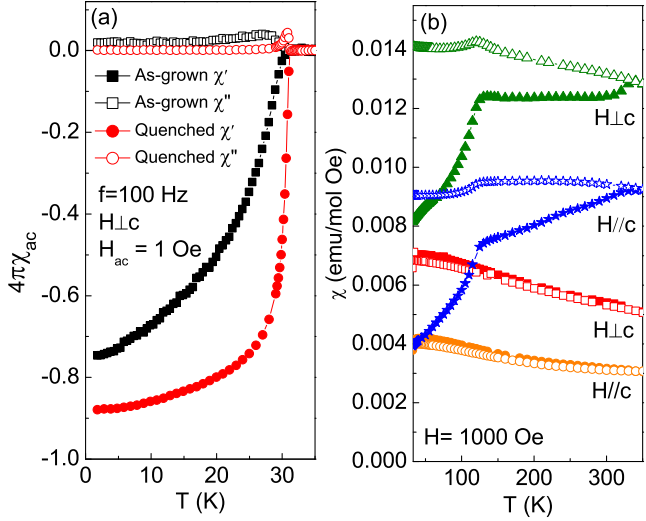}} \vspace*{-0.3cm}
\caption{(a) Temperature dependence of ac magnetic susceptibility for
as-grown (squares) and quenched (circles) K$_{0.69(2)}$Fe$_{1.45(1)}$Se$%
_{2.00(1)}$ taken in H=1 Oe. (b) Temperature dependence ZFC (filled symbols)
and FC (open symbols) dc magnetic susceptibility for as-grown (squares and
circles) and quenched (triangles and stars) K$_{0.69(2)}$Fe$_{1.45(1)}$Se$%
_{2.00(1)}$ in H=1000 Oe.}
\end{figure}

The superconducting volume fractions at 1.8 K of K$_{0.69(2)}$Fe$_{1.45(1)}$%
Se$_{2.00(1)}$ crystals increased by annealing and quenching process from $%
\sim$ 74.6\% to $\sim$ 87.9\% (Fig. 1(a)). Moreover, the superconductivity
in quenched sample is more homogeneous and sharper at T$_{c}$ $\sim $ 30K
than in as-grown samples (Fig. 1(a)). In addition, there is a significant
enhancement of susceptibility in the normal state after post-annealing and
quenching process (Fig. 1(b)).

\begin{figure}[tbp]
\centerline{\includegraphics[scale=1.6]{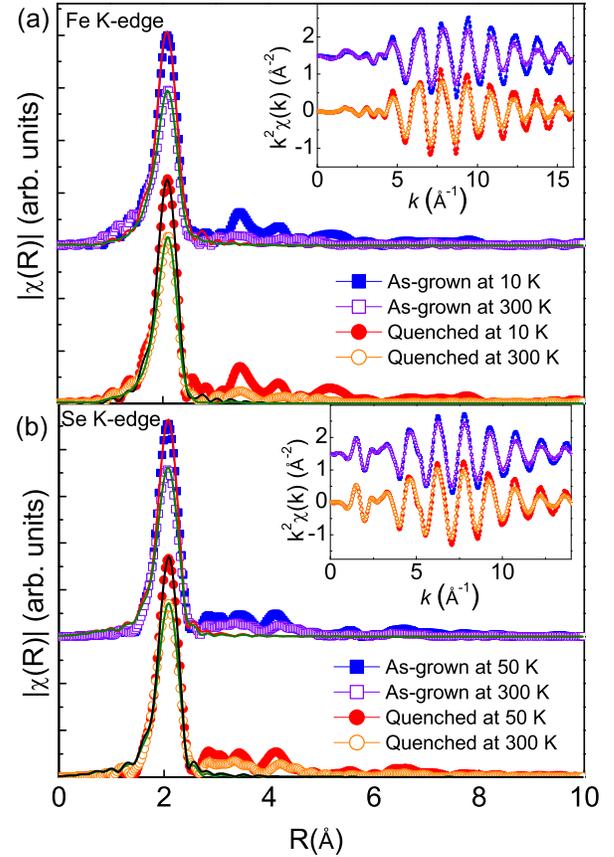}} \vspace*{-0.3cm}
\caption{Representative Fourier transform (FT) magnitudes of EXAFS data. Fe
K-edge results at 10 K and 300 K of as-grown and quenched K$_{0.69(2)}$Fe$%
_{1.45(1)}$Se$_{2.00(1)}$ samples are shown in (a), and Se K-edge results at
50 K and 300 K of both samples are shown in (b). Corresponding EXAFS
oscillations are shown in the insets. The FTs are representing raw
experimental data without correcting for the phase shifts. The theoretical fits are shown as solid lines.}
\end{figure}

The first nearest neighbors of Fe atoms are Se atoms located at about 2.4
\AA\ distance, and the second nearest neighbors of Fe atoms are Fe atoms, at
about 2.8 \AA . The first nearest neighbors of Se atoms are Fe atoms with
bond distances around 2.4 \AA , and the second nearest neighbors are Se
atoms, at about 3.9 \AA\ . The peaks around 2 \AA\ (Fig. 2) correspond to the Fe-Se and Fe-Fe bond distances (the peak positions are not corrected for the photoelectron phase shifts) for Fe K-edge data (Fig. 2(a)) and only to the Fe-Se bond distances for the Se K-edge data (Fig. 2(b)). The actual distance values were
extracted from the theoretical fits.

K$_{0.69(2)}$Fe$_{1.45(1)}$Se$_{2.00(1)}$ has two different Fe sites, Fe1
and Fe2. High symmetry Fe1 site has four Se as first nearest neighbors with
identical Fe-Se distance, 2.4850(12) \AA\ determined by the average
structure.\cite{Hechang} In contrast, lower symmetry Fe2 site has four
nearest neighbor Se atoms with four different Fe-Se bond lengths, 2.3956(12)
\AA , 2.4632(13) \AA , 2.4061(12) \AA , and 2.4923(13) \AA .\cite{Hechang}
Since XAFS probes all Fe sites, the ratio of higher symmetric site (Fe1) and
lower symmetric site (Fe2) occupancies can be determined by the average
theoretical Fe-Se bond length behavior if both sites are occupied. The
average Fe-Se bond length as obtained by XRD is 2.44 \AA\ (2.48 \AA ) when only lower (higher)
symmetric site is occupied. Therefore, the average Fe-Se bond length will increase with the increased occupancy of the high symmetry site.
The opposite trend is expected for the static bond length disorder. It is expected to be 0.0016 \AA $^{2}$ with only low symmetry (Fe2) site is occupied, whereas it is 0 when only high symmetry (Fe1) is occupied. Hence, the disorder values should decrease with the increased occupancy of the high symmetry site.

\begin{figure}[tbp]
\centerline{\includegraphics[scale=1.05]{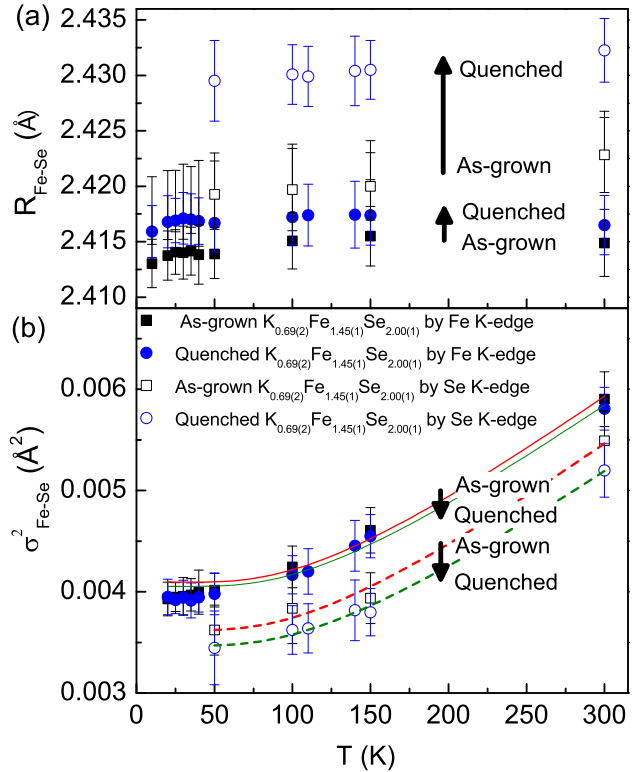}} \vspace*{-0.3cm}
\caption{(a) Temperature dependence of the Fe-Se distances obtained from the
Fe K-edge (filled symbols) and Se K-edge (open symbols) for as-grown
(squares) and quenched (circles) K$_{0.69(2)}$Fe$_{1.45(1)}$Se$_{2.00(1)}$.
(b) Mean square relative displacements $\protect\sigma ^{2}$ for the nearest
neighbor Fe-Se shell derived from Fe K-edge analysis (filled symbols) and Se
K-edge analysis (open symbols) for as-grown (squares) and quenched (circles)
K$_{0.69(2)}$Fe$_{1.45(1)}$Se$_{2.00(1)}$. $\protect\sigma_{s}^{2}$\
decreases after quenching indicating more uniform nearest-neighbor Fe-Se
shell after quenching. The arrows show the trend of change after post-annealing and quenching process.}
\end{figure}

The XAFS bond distance values are smaller than the Rietveld values. We note that the Fe-Se bond distance was measured from either Fe or Se edge by the Gaussian
approximation for the distinct non-Gaussian bond length distribution. Most
of the Fe-Se bond distances are distributed on the large distance side,
including bond lengths with 2.4632(13) \AA , 2.4850(12) \AA , and 2.4923(13)
\AA . Nevertheless few still remain on the lower distance side, thus biasing
(lowering) the Gaussian peak position. The Se K-edge Fe-Se bond distances appear larger when compared to the Fe K-edge bond distance (Fig. 3(a)). Even
though the XAFS bond distances are smaller when compared to Rietveld values,
the relative change in static disorder extracted from edge is still a reliable measure of relative structural changes.
The Fe-Se bond distances increase after quenching (Fig. 3(a)). This is consistent with the
expected result that Fe and Se nearest neighbor distance is supposed to increase as Fe1 occupancy increases.
In what follows we will focus on the relative change in the occupancies between Fe1 and Fe2 sites.

The mean square relative displacements (MSRD) describe distance-distance
correlation function (correlated Debye-Waller factors). They include
contributions from the temperature-independent term, $\sigma _{s}^{2}$, and
the temperature-dependent term, $\sigma _{d}^{2}(T)$, i.e. $\sigma ^{2}=$ $%
\sigma _{s}^{2}+\sigma _{d}^{2}(T)$.\cite{Prins} The subscripts \emph{s} and \emph{d} mean static and dynamic, respectively. Temperature-dependent term is
well described by Einstein model:\cite{Prins}
\begin{equation}
\sigma _{d}^{2}(T)=\frac{\hbar }{2\mu \omega _{E}}\coth (\frac{\hbar \omega _{E}%
}{2k_{B}T}),\newline
\end{equation}%
where $\mu $ is the reduced mass of the Fe-Se bond and $\omega _{E}$ is the
Einstein frequency related to the Einstein temperature $\theta _{E}=\hbar
\omega _{E}/k_{B}$. The fitting curves for as-grown K$_{0.69(2)}$Fe$%
_{1.45(1)}$Se$_{2.00(1)}$ Fe K-edge (red solid line) and Se K-edge (red
dotted line) give $\theta _{E}$ =(353 $\pm $ 22) K and $\theta _{E}$ = (355 $%
\pm $ 10) K, respectively. Similar analysis for the quenched sample yields $%
\theta _{E}$ = (359 $\pm $ 19) K and $\theta _{E}$ = (364 $\pm $ 4) K for Fe
K-edge and Se K-edge, respectively. The results are identical within error
bars. The relative difference between as-grown and quenched sample static
disorder points to the possible rearrangement of Fe1 and Fe2 site
occupancies. The static disorder $\sigma _{s}^{2}$ values obtained from the
fits are 0.0020 $\pm $\ 0.0002 \AA $^{2}$ for Fe K-edge of both as-grown and
quenched samples, and 0.00150 $\pm $\ 0.00012 \AA $^{2}$ and 0.00140 $\pm $\
0.00004 \AA $^{2}$ for Se K-edge of as-grown and quenched samples,
respectively. The local force constant
k can be calculated from $k=\mu \omega _{E}^{2}$.\cite{Joseph} For as-grown
and quenched K$_{0.69(2)}$Fe$_{1.45(1)}$Se$_{2.00(1)}$ local force constants
of Fe-Se bonds are 7.32 $\pm $\ 0.29 eV/\AA $^{2}$and 7.70 $\pm $\ 0.12 eV/%
\AA $^{2}$, respectively, indicating that the Fe-Se bond hardens after
quenching. This is consistent with higher degree of bond order.

Experimentally measured behaviors of the Fe-Se distance (increases in the quenched sample) and its static disorder (decreases in the quenched sample) are consistent with the trends described above and thus can be attributed to the increased occupancy of the high symmetry site in the quenched sample (Fig. 3(a) and (b)). Also, the magnetic moment
on quenched K$_{0.69(2)}$Fe$_{1.45(1)}$Se$_{2.00(1)}$ samples doubled (Fig.
1(b)). These results provide clear evidence that Fe1 sites can be associated
with much higher magnetic moment (more than five times larger) than Fe2
sites.\cite{Zavalij}

What are the implications of our results on nanoscale phase separation and
vacancy disordered superconducting phase in K$_{0.69(2)}$Fe$_{1.45(1)}$Se$%
_{2.00(1)}$?\cite{Wang,Ricci,Ricci1} Superconducting K$_{x}$Fe$_{2-y}$Se$%
_{2} $ crystals are found for a partially broken iron vacancy order,\cite%
{Yan} corresponding to narrow region of Fe valence from 2 to about 1.94.
This corresponds to deviation from ideal K$_{0.8}$Fe$_{1.6}$Se$_{2.00(1)}$
stoichiometry (or more general from K$_{1-x}$Fe$_{1.5+(x/2)}$Se$_{2}$) where
Fe1 site is empty and Fe2 is completely occupied (K$_{2}$Fe$_{4}$Se$_{5}$
phase).\cite{Yan,Ricci1,Song} Thus, superconducting crystals are found for K$%
_{0.8}$ and excess Fe content Fe$_{1.6+x}$ (x$\>>$0)\cite{Wang,Yan}
suggesting broken vacancy order by some finite Fe1 occupancy, or for K$%
_{0.69(2)}$ and Fe$_{1.45(1)}$ stoichiometry suggesting broken vacancy order
by deficiency on both K and Fe2 sites.\cite{Hechang} This is in agreement
that K content is rather important for superconductivity.\cite{Yan} Since
nominal stoichiometry in our as-grown superconducting crystals was K$%
_{0.69(2)}$Fe$_{1.45(1)}$Se$_{2.00(1)}$ with only Fe2 site occupied,\cite%
{Hechang} the increased occupancy of Fe1 sites in quenched
crystals implies further depletion of Fe2 sites and stronger deviation from
vacancy ordered K$_{0.8}$Fe$_{1.6}$Se$_{2.00(1)}$ insulating phase.\cite%
{Wang} Our results provide the first structural evidence that local structure disorder
and Fe site occupancy in K$_{3}$Fe$_{4}$Se$_{5}$ is the key structure factor
for bulk and homogeneous superconductivity in high-T$_{c}$ iron based
superconductor K$_{x}$Fe$_{2-y}$Se$_{2}$.

\section{Conclusion}

In summary, the temperature dependent XAFS study of Fe and Se \textit{K}%
-edge spectra of as-grown and quenched K$_{0.69(2)}$Fe$_{1.45(1)}$Se$%
_{2.00(1)}$ samples indicates that the average Fe-Se bond distance increases
after quenching due to the increase in population of high symmetry Fe1 sites
which have higher bond distance. For both samples, the temperature
dependence of the MSRD of the Fe-Se bonds follows the Einstein model. The
static disorder results ($\sigma _{s}^{2}$) show that the atoms are more
ordered after post-annealing and quenching process, also pointing to the
increase in Fe1 high symmetry site. Finally, based on the local force
constant analysis, Fe-Se bonds become stronger after post-annealing and
quenching. This is consistent with the above analysis. Increased occupancy
of high symmetry Fe1 site coincides with the large increase in paramagnetic
moment, indicating that Fe1 sites are strongly magnetic. Simultaneously and surprisingly, superconductivity volume fraction is
increased and superconducting T$_{c}$ is much sharper, suggesting more
homogeneous superconducting crystal. Our results testify that
superconducting volume fraction and homogeneity of superconducting phase is
in direct competition with Fe vacancy order.

\begin{acknowledgements}

We thank John Warren for help with scanning electron microscopy measurements
and S. Khalid for help with XAFS measurements. Work at Brookhaven is
supported by the U.S. DOE under Contract No. DE-AC02- 98CH10886 and in part
by the Center for Emergent Superconductivity, an Energy Frontier Research
Center funded by the U.S. DOE, Office for Basic Energy Science (H. L. and C. P.). A.I.F.
acknowledges support by U.S. Department of Energy Grant DE-FG02-03ER15476.
Beamline X19A at the NSLS is supported in part by the U.S. Department of
Energy Grant No DE-FG02-05ER15688.

\end{acknowledgements}

\end{document}